\pdfoutput=1
\documentclass[11pt]{article}

\usepackage[utf8]{inputenc}
\usepackage[T1]{fontenc}
\usepackage[margin=1in]{geometry}
\usepackage{hyperref}
\usepackage{url}
\usepackage{booktabs}
\usepackage{amsfonts}
\usepackage{nicefrac}
\usepackage{microtype}
\usepackage{xcolor}
\usepackage{tabularx}

\hypersetup{
  colorlinks=true,
  linkcolor=black,
  citecolor=black,
  urlcolor=blue
}

\title{Urban AI Governance Must Embed Legal Reasonableness for Democratic and Sustainable Cities}

\author{Rashid Mushkani\\
Université de Montréal\\
Mila -- Quebec AI Institute}

\date{}

\begin{document}
\maketitle

\begin{abstract}
\textbf{This position paper argues that embedding the legal ``reasonable person'' standard in municipal AI systems is essential for democratic and sustainable urban governance.} As cities increasingly deploy artificial intelligence (AI) systems, concerns around equity, accountability, and normative legitimacy are
growing. This paper introduces the Urban Reasonableness Layer (URL), a
conceptual framework that adapts the legal ``reasonable person''
standard for supervisory oversight in municipal AI systems, including
potential future implementations of Artificial General Intelligence
(AGI). Drawing on historical analogies, scenario mapping, and
participatory norm-setting, we explore how legal and community-derived
standards can inform AI decision-making in urban contexts. Rather than
prescribing a fixed solution, the URL is proposed as an exploratory
architecture for negotiating contested values, aligning automation with
democratic processes, and interrogating the limits of technical
alignment. Our key contributions include: (1) articulating the
conceptual and operational architecture of the URL; (2) specifying
participatory mechanisms for dynamic normative threshold-setting; (3)
presenting a comparative scenario analysis of governance trajectories;
and (4) outlining evaluation metrics and limitations. This work
contributes to ongoing debates on urban AI governance by foregrounding
pluralism, contestability, and the inherently political nature of
socio-technical systems.
\end{abstract}

\section{Introduction}

Urban design has long mediated tensions between formal efficiency and
socio-political meaning. Sullivan's dictum that ``form follows
function'' captured a modernist faith in optimization, yet subsequent
scholarship has demonstrated that artefacts, whether buildings or
algorithms, always embed political choices (Winner, 1980). The diffusion
of information and communication technologies extended built form into
what Floridi terms the ``infosphere,'' a space where physical
structures, data streams, and social expectations are co-constitutive
(Floridi, 1999; 2002).

\textbf{This paper takes the position that urban AI governance must embed a dynamic, democratically grounded standard of legal reasonableness to align automated decision-making with pluralistic values and sustainability.}

Projections suggest that by 2030, automation could account for up to 30
percent of work hours in the United States, with generative AI
accelerating these trends. Approximately 60 percent of occupations
include tasks in urban sectors such as transport and supervision that
are susceptible to partial automation (Ellingrud et al., 2023; McKinsey
Global Institute, 2017; Korinek, 2023). This trajectory raises questions
about the evolving societal purpose of cities, which have historically
served as centers of employment and consumption. Simultaneously, policy
analyses caution that ungoverned AI deployment may intensify
surveillance, exacerbate socioeconomic inequalities, and centralize
decision-making power within opaque algorithmic systems (Attard-Frost,
2025; European Commission, 2018; Gebru \& Torres, 2024).

Legal and policy responses remain fragmented. Rights-based regulatory
proposals often promise broad protections yet provide limited
justiciability once systems are deployed (Mei \& Robitscher, 2025).
Competing ``digital empires'' of the United States, the European Union,
and China export divergent regulatory logics that shape global
technology supply chains (Bradford, 2023). Meanwhile, philosophical
analyses warn that assigning either full personhood to AI (Solum, 1992)
or intrinsic value to informational artefacts (Floridi, 1999) demands
careful integration with existing human rights regimes. Recent
scholarship has expanded these debates to include emerging claims to a
participatory "right to AI," which conceptualizes AI as a societal
infrastructure requiring collective governance and data stewardship
(Mushkani et al., 2025a), alongside broader propositions for a "right to
technology" (Sun, 2020). Empirical work on moral consideration for
artificial entities further underscores the plurality of public
intuitions that any urban governance framework must address (Harris \&
Anthis, 2021).

At its core, the URL draws on the Anglo-American “reasonable person” standard, a foundational legal doctrine used to evaluate conduct in negligence cases since Vaughan v. Menlove (1837) and Blyth v. Birmingham Waterworks (1856). Originally conceived as an objective measure of prudent behavior, the test was refined in United States v. Carroll Towing (1947) through Judge Learned Hand’s cost-benefit calculus, and later extended into administrative and constitutional law to assess whether governmental actions proportionately balance competing interests (Farnsworth, 1967; Schwartz, 1989; Wright, 2002). Contemporary legal theory increasingly treats reasonableness as a socially em-bedded threshold, responsive to shifts in public norms, technological capacity, and acceptable risk. Emerging work on algorithmic accountability has begun to adapt this doc-trine for machine behavior, proposing that dynamically up-dated profiles of what a well-informed, fair-minded citizen would consider reasonable can support transparency and democratic oversight in AI systems (Rane, 2024). Framing reasonableness in this way positions it not as a static moral truth but as a flexible evaluative lens, one that can help align AI decision-making with evolving social expectations and civic legitimacy.

This paper seeks to address three interrelated questions: How can the
legal standard of the ``reasonable person'' be operationalized as a
supervisory tool within municipal AI and AGI systems? What participatory
mechanisms and technical infrastructures are required to ensure these
systems reflect democratic and pluralistic values? And what risks must
be mitigated to prevent the deepening of opacity and inequity as AI
becomes foundational to urban governance?

Recent advances in AI research outline both potential pathways and
limitations. Zhang et al. (2025) propose the Autonomous Generalist
Scientist to illustrate progression from tools to self-directed agents,
yet empirical analyses indicate that apparent AGI capabilities often
emerge from statistical correlations rather than genuine normative
understanding (Altmeyer et al., 2024). Dai (2024) distinguishes
mechanistic paradigms, which reduce agency to input--output
optimization, from volitional paradigms that emphasize intrinsic purpose
and normative judgment, qualities absent in current AI systems.
Concurrently, regulatory frameworks struggle to align ambitious mandates
with nascent technical capabilities, risking either unenforceable
standards or constrained innovation (Maslej et al., 2024; Trager et al.,
2023).

Despite extensive work on smart cities and technical alignment, urban
scholarship has yet to integrate these insights within democratic and
environmental agendas comprehensively. This paper addresses that gap by
extending historical analyses of prosthetic subjectivity in ancient Rome
(Morley, 2017; Deibel \& Deibel, 2023), refining conceptual distinctions
between AI urbanism and legacy smart city paradigms, introducing the URL
as a supervisory mechanism embedding the legal ``reasonable person''
standard in municipal AI and AGI systems, and proposing policy-relevant
design principles and a scenario matrix to guide equitable and
sustainable AI deployment.

To clarify both the novelty and applied potential of this work, the
paper advances the state of the art by detailing how the URL can be
instantiated in actual municipal contexts. This includes a comparative
analysis with existing smart city governance models, a breakdown of the
technical and organizational steps necessary for participatory
threshold-setting, and identification of audit and contestation
mechanisms. By prioritizing operational specificity, the proposed
framework bridges conceptual insights with real-world implementation,
enhancing its value to both academic and policy audiences.

This paper further recognizes that any framework for urban AI governance
must center the perspectives of historically marginalized communities,
including racial minorities and people with disabilities. Drawing from
decolonial and critical disability studies, it argues for their
systematic inclusion in the participatory processes of the URL to
prevent the perpetuation of structural injustices in emerging urban
systems.

\section{Method}
This study combined interpretive analysis with socio-technical
transition design, drawing on primary sources including classical texts,
contemporary policy white papers, corporate technical documents, and
peer-reviewed scholarship. We began by periodizing successive waves of
labor substitution, from enslaved ``living tools'' to mechanization,
electrification, digital automation, and AI. Each stage was analyzed
along three dimensions: the distribution of agency, the velocity of
diffusion, and the spatial impacts on urban environments.

Building on this foundation, socio-technical systems were mapped by
linking material artifacts (such as sensors and actuators),
institutional regimes (laws, norms), and interpretive frames (public
narratives), following the multi-level perspective outlined by Geels
(2002). Participatory capacity was prioritized using Arnstein's ladder
of citizen participation and Lefebvre's ``right to the city'' to
foreground the political economy of inclusion (Arnstein, 1969; Lefebvre,
1968).

To explore future possibilities, we constructed scenarios by varying
uncertainties around ownership (public, private, commons), governance
models (participatory, technocratic, authoritarian), and civic agency to
derive coherent urban futures. Throughout the process, five expert peer
debriefings were conducted to enhance reflexivity and ethical rigor.
Participants included specialists in AI ethics, urban planning, law, and
municipal governance, including staff from the City of Montreal. These
sessions aligned with best practices in emerging-technology assessment
(Harris \& Anthis, 2021). Rather than predict outcomes, the aim was to
articulate plausible pathways to guide future empirical research and
policy development.

\section{Historical Analogies}
Historical frameworks for labor and technology provide important context
for understanding contemporary debates in urban AI governance. Classical
sources describe Roman slaves as ``living tools,'' a concept that
institutionalized automation as both an economic and social structure
(Morley, 2017). Aristotle's classification of slaves as ``instruments of
action,'' examined through subsequent archaeological analysis,
demonstrates how technological metaphors have long served to rationalize
social hierarchies (Morley, 2017). These precedents resonate with
ongoing discussions about the legal status of artificial agents,
specifically whether AI systems should be accorded legal personhood or
treated as property (Solum, 1992; Gunkel, 2006).

Current forms of urban AI and the prospective development of AGI diverge
from previous technological shifts in several respects. Contemporary AI
systems are designed primarily to automate specific tasks or to provide
analytic support within bounded municipal domains. By contrast,
envisioned AGI systems are characterized by the goal of domain-general
coordination, integrating functions across logistics, finance, and
governance. The adoption of cloud infrastructure by current AI
applications already accelerates global diffusion and reduces
infrastructural adaptation cycles, reshaping established patterns of
urban transformation (Korinek \& Suh, 2023; Kim et al., 2024). The
potential future emergence of AGI introduces additional considerations,
such as the capacity for recursive self-improvement, which may lead to
capability escalation and systemic lock-in as theorized in
superintelligence scenarios (Bostrom, 2014; Gebru \& Torres, 2024).

These developments often align with ideological currents collectively
described by Gebru and Torres (2024) as the TESCREAL bundle, an acronym
for transhumanism, Extropianism, singularitarianism, (modern) cosmism,
Rationalism, Effective Altruism, and longtermism. They critique this
cluster for promoting a utopian vision that privileges speculative
futures over present-day political and ethical concerns.

To structure the allocation of autonomy in municipal AI systems, Zhang
et al. (2025) propose distinct tiers: tool, assistant, associate, and
pioneer. Mapping municipal functions onto these levels makes explicit
the trade-offs between efficiency and oversight. As Winner (1980)
observes, the long-term distribution of power in technological systems
is determined less by technical design than by the institutional and
legal regimes in which those systems are embedded. Therefore, embedding
normative alignment mechanisms such as the URL at the design stage is
essential for ensuring accountability and democratic legitimacy in urban
AI governance.

\section{Intelligent Cities}
Traditional smart-city programs augment human decision-makers with
narrow AI models and dense sensor networks (Table 1 compares Smart-City
and AGI-driven city models). However, as Winner (1980) argues, technical
infrastructures are never value-neutral. These systems typically support
discrete subsystems, such as traffic management, utilities, and waste
collection, offer analytics rather than autonomous decision-making,
undergo periodic retraining, and operate under vendor contracts and
municipal IT oversight.

Comparative frameworks highlight key differences between smart cities and AGI-driven city models (Table~\ref{tab:smart-agi}).

\begin{table}[ht]
\centering
\caption{Comparative dimensions of smart-city and AGI-driven city models.}
\label{tab:smart-agi}
\begin{tabularx}{\textwidth}{p{0.18\textwidth} p{0.36\textwidth} X}
\toprule
\textbf{Dimension} & \textbf{Smart City} & \textbf{AGI-Driven City} \\
\midrule
Agency     & Analytics support human officials               & AGI proposes, executes, self-optimizes \\
Scope      & Discrete subsystems (traffic, utilities)        & Integrated system-of-systems optimization \\
Learning   & Periodic manual retraining                      & Continuous, self-directed adaptation \\
Governance & Vendor contracts, municipal IT units            & Layered oversight, constitutional limits, citizen veto \\
\bottomrule
\end{tabularx}
\end{table}

In contrast, an AGI-driven city would generate causal narratives,
self-optimize across interdependent subsystems, and adapt to novel
conditions through continuous, self-directed learning (Bostrom, 2014;
Kim et al., 2024). Governance in this paradigm would shift from fixed
contractual arrangements to layered oversight architectures,
incorporating constitutional limits and explicit citizen-veto mechanisms
(Bradford, 2023).

Adjustable-autonomy theory (Mostafa et al., 2019; Pynadath et al., 2002)
provides a framework for mapping human--machine complementarities,
spanning systems that deliver metrics to those that propose novel
policies subject to strategic human review. Table 2 presents autonomy
tiers applicable to urban AGI governance. Mapping municipal functions,
such as energy dispatch, zoning decisions, and flood response, onto this
framework illustrates how varying autonomy levels can balance adaptive
efficiency with normative oversight. For example, Shanghai's AI-based
traffic control system reduced peak-period congestion by approximately
fifteen percent (Zhang et al., 2019), illustrates benefits at
intermediate autonomy levels. Early deployments of AGI systems in urban
governance should therefore initiate at lower autonomy tiers, advancing
incrementally following iterative calibration of URL thresholds and
sustained public engagement to legitimate expansions of decision-making
authority (Arnstein, 1969). Adjustable-autonomy theory, as discussed by Mostafa et al.\ (2019) and Pynadath et al.\ (2002), provides a framework for mapping human–machine complementarities (Table~\ref{tab:autonomy}).

\begin{table}[ht]
\centering
\caption{Adjustable autonomy tiers for urban AGI governance.}
\label{tab:autonomy}
\begin{tabularx}{\textwidth}{p{0.08\textwidth} p{0.22\textwidth} X}
\toprule
\textbf{Tier} & \textbf{Role} & \textbf{Description} \\
\midrule
0 & Tool & Human designs and executes; AI supplies performance metrics. \\
1 & Assistant & AI proposes options; human selects among alternatives. \\
2 & Associate & AI executes routine tasks; human intervenes on exceptions. \\
3 & Pioneer & AI explores novel policies; humans review strategic outcomes. \\
\bottomrule
\end{tabularx}
\end{table}

\section{The Urban Reasonableness Layer (URL)}

Technical-alignment research often neglects context-specific normative
benchmarks, contributing to what some scholars characterize as the
"illusion" of rights-based AI governance (Mei and Robitscher, 2025;
Bradford, 2023). Rane et al. (2024) propose adapting the legal
``reasonable person'' standard into computational assessments. In tort
and contract law, reasonableness mediates responsibility by evaluating
actions against community norms (Farnsworth, 1967; Schwartz, 1989;
Wright, 2002). The URL implements this standard as a supervisory control
structure integrated into AI systems and potential future AGI
decision-making modules.

Defining reasonable thresholds within the URL should itself be subject
to participatory processes, comprising several months of open
deliberation on accessible digital platforms. Citizens would engage via
forums, asynchronous discussion threads, and multimedia contributions,
supported by audio descriptions, screen-reader interfaces, and
multilingual transcripts, to accommodate diverse forms of
representation, including persons with visual impairments. Over
iterative cycles, participants would propose, assess, and refine
normative metrics and thresholds, thereby establishing a
community-endorsed corpus of interpretive standards. The resulting
dataset of deliberated norms would serve as the reference for the URL's
assessor module, ensuring that algorithmic judgments reflect collective
expectations rather than solely expert judgments (Arnstein, 1969;
Mushkani et al., 2025b).

The URL architecture consists of four interconnected stages. First,
input parsing transforms natural-language prompts and multisensor data
into structured intent vectors. Second, a context-retrieval engine
identifies relevant bylaws, legal precedents, and participatory metrics.
Third, an assessor module applies reasonableness tests, issuing penalty
signals for detected deviations. Fourth, a reinforcement-update
mechanism incorporates assessor feedback into the system's reward
functions, shaping subsequent policy exploration.

Implementing the URL raises several challenges. Jurisdictional
variability introduces contextual ambiguities; hierarchical referencing
across local, national, and international norms mitigates this risk,
while participatory audits periodically recalibrate thresholds to
reflect evolving community standards. For instance, if a flood-response
agent proposes rerouting traffic through disadvantaged neighborhoods,
the URL flags disproportionate burdens and prompts iterative adjustments
until equity metrics are satisfied.

For the URL framework to have meaningful impact, clear evaluation
metrics are essential. These metrics should encompass procedural
indicators, such as participation rates, diversity of deliberative
input, and transparency of threshold adjustments, as well as substantive
outcomes, including reductions in algorithmic bias, improved
distributive equity, and public satisfaction with automated decisions.
Technical benchmarks may involve system robustness to alignment drift,
auditability of decision logs, and frequency of successful citizen
overrides. Systematic monitoring of these indicators will enable both
iterative improvement and comparative evaluation against baseline
governance models.

The technical implementation of the URL requires integration with
municipal data infrastructures, development of secure digital
participation platforms, and design of explainable AI modules capable of
surfacing and justifying reasonableness assessments. A modular pilot
program, beginning with a single domain, such as traffic routing or
public health resource allocation, would allow for incremental learning
and risk mitigation. Partnerships with municipal departments and local
advocacy groups will be essential to ensure legitimacy and iterative
improvement.

Anchoring the reward functions of AI systems, and of prospective AGI
systems as they emerge, in jurisprudential and participatory data
addresses two primary critiques. First, emergent model representations
in AI may capture spurious correlations rather than reflecting normative
validity (Alain \& Bengio, 2016; Gurnee \& Tegmark, 2023). Second, many
current approaches to technical alignment lack domain-specific
benchmarks and systematic safety diagnostics, a limitation that may
become more pronounced as AGI capabilities advance (Sorensen et al.,
2024; Christian, 2020; Gebru \& Torres, 2024; Vidgen et al., 2023). By
incorporating both legal standards and community-derived norms into
supervisory control architectures, the URL maintains a continuous
alignment between algorithmic decision-making and evolving societal
expectations. This configuration allows the URL to connect mechanistic
control with volitional oversight by embedding participatory and
jurisprudential processes directly within the governance structures of
present-day AI and future AGI systems.

\section{Infrastructure, Design, and Environmental Performance}
Urban infrastructure operates within the context of significant network
externalities and protracted capital life cycles, which shape the
capacity for adaptation and renewal (Batty, 2018; Zhang et al., 2019).
While current generative models are limited to generating static
schematic plans under predetermined constraints (Sanchez et al., 2024;
Ramesh et al., 2021), prospective AGI systems, when supervised by
frameworks such as the URL, may perform
iterative design simulations, assess socio-environmental impacts, and
generate adaptive urban morphologies by leveraging large-scale
transformer architectures (Batty, 2018; Petrov et al., 2024).

The implementation of URL-supervised AGI within urban infrastructure
requires effective cross-sector coordination. Aligning incentives among
public agencies, private vendors, and affected communities depends on
transparent communication strategies and inclusive stakeholder mapping.
Such coordination is essential to prevent the disproportionate capture
of benefits by specific interest groups.

The integration of normative trade-offs, such as reconciling carbon
neutrality objectives with heritage preservation, depends on the
collaborative design of evaluation metrics. This process requires
substantive engagement with impacted communities to prevent the
emergence of algorithmic monocultures (Mushkani et al., 2025a). For
example, integrated energy--transport systems that combine real-time
demand management with vehicle-to-grid balancing have been projected to
yield significant reductions in CO$_2$ emissions (Bibri \& Krogstie, 2019).
However, the open-ended learning dynamics characteristic of advanced AI systems may still generate unanticipated system behaviors, underscoring
the necessity for superalignment protocols and regular expert audits to
maintain alignment with community standards (Kim et al., 2024).

Embedding URL-supervised AGI within ecological feedback systems enables
the real-time operationalization of environmental objectives. For
instance, dynamic penalty signals can be issued when carbon indicators
exceed community-established thresholds, facilitating adaptive
governance that maintains operational continuity while advancing
environmental goals (Floridi, 2002). This configuration supports the
alignment of optimization processes with environmental values that have
been explicitly articulated through participatory mechanisms.

\section{Political Economy of the Post‑Work City}
The progression toward AGI-driven automation is transforming the
economic and institutional foundations of urban environments by
detaching labor from production and altering the distribution of value
(Acemoglu \& Restrepo, 2018; Stiefenhofer, 2025). As automation reduces
marginal labor costs, the relative bargaining power of human workers
diminishes, contributing to wage compression and the concentration of
income and assets among capital owners. These shifts present structural
challenges that extend beyond the labor market, necessitating a
reevaluation of municipal finance, urban spatial organization, and
collective civic agency.

Conventional municipal finance relies on payroll taxes, commercial
rents, and transit fares, all of which are susceptible to decline under
conditions of widespread automation and remote operation facilitated by
cloud-hosted AGI. This dynamic creates fiscal pressures for local
governments and compels consideration of new policy instruments. Among
these are AGI-compute royalties, progressive taxation on robotic
capital, and municipal wealth funds financed through licensing or access
fees (Bradford, 2023; Korinek \& Suh, 2023). Site-value taxation, which
channels location-based rents into universal basic dividends and
supports the provision of public goods, represents an additional
approach to stabilizing municipal budgets as wage-derived revenues
recede (Mei \& Robitscher, 2025).

The decoupling of labor from place further introduces the potential for
renewed polycentricity and rural regeneration. While productive activity
becomes increasingly untethered from fixed locations, urban amenity
cores may continue to function as sites of social interaction and
exchange, thereby maintaining rent gradients and mitigating
congestion-related externalities. Regional development strategies that
align technological infrastructure with localized objectives can serve
as counterweights to the concentration of economic activity and
population in metropolitan centers.

The displacement of waged work also has implications for social identity
and civic participation (Arendt, 1998). As traditional employment
declines in significance, risks arise for collective belonging and
political engagement. Investments in educational initiatives, cultural
programs, and participatory governance, situated in civic institutions
such as libraries and makerspaces, may cultivate new forms of motivation
and agency. Mechanisms that direct automation-derived revenues into
participatory budgeting or deliberative processes align the distribution
of technological gains with social cohesion and democratic governance
(Arnstein, 1969; Lefebvre, 1968; Mushkani et al., 2025a).

The URL operates as an institutional mechanism for mediating normative
trade-offs in the post-work city. By embedding participatory and legal
standards within supervisory control of AGI, the URL enables communities
to shape the criteria by which resource allocation and urban planning
decisions are made. This model addresses the risk that the gains from
automation may reinforce existing patterns of inequity in the absence of
intentional and inclusive governance.

\section{Governance}
The structuring of governance architectures in urban AI deployment is
shaped by the underlying definition of ``the public.'' If individuals
are framed primarily as consumers or data subjects, there is a risk of
systematically excluding marginalized voices from deliberative and
supervisory processes (Sieber et al., 2024; Cardullo, 2020).
Furthermore, the institutional challenge of recognizing both human and
artificial actors complicates the design of effective oversight
structures (Solum, 1992).

Several governance archetypes can be identified. The AI-assisted
bureaucracy retains elected authority while assigning AGI systems the
function of preparing policy options for human review. Algorithmic
technocracy with democratic override restricts AGI operation within
constitutional boundaries and assigns explicit veto rights to citizens
(Floridi, 1999; Wenar, 2023). Trans-jurisdictional supervision licenses
local AGI instances subject to independent safety audits modeled on
frameworks such as that of the International Atomic Energy Agency
(OpenAI, 2023). However, such arrangements may create a perception of
rights-based regulation without corresponding enforcement capacity,
thereby limiting their practical effectiveness (Mei \& Robitscher,
2025).

Exclusive reliance on technical governance solutions risks entrenching
``algorithmic managerialism'' and may limit the adaptability of
democratic processes. More robust approaches layer constitutional
protections, municipal charters, and participatory mechanisms, producing
polycentric and adaptive governance structures capable of continuous
recalibration as technological capacities and civic expectations evolve.

Establishing and maintaining trust in urban AI governance depends on the
availability of transparent and contestable processes. Public dashboards
that disclose system optimization targets, performance metrics, and
instances of manual override are instrumental in supporting civic
confidence (Castelfranchi \& Falcone, 2000; Vidgen et al., 2023).
Continuous red-team testing and open publication of safety benchmarks
are necessary to address recognized failure modes (Amodei et al., 2016).
Since technical artifacts embody political choices (Winner, 1980;
Gunkel, 2018), regular foresight exercises (including scenario
workshops, deliberative polling, and digital twin simulations) are
required to support the ongoing recalibration of governance protocols in
accordance with evolving societal values (Taylor, 1985).

\section{Risks and Scenarios}
Automated urban governance systems generate intersecting risk categories
that require systematic mitigation. Alignment drift, in which the
objectives specified for AI systems diverge from actual system
behaviors, can be addressed through continuous audits of URL,
superalignment protocols, and robustness assessments of representational
models (Kim et al., 2024; Qi et al., 2023; Petrov et al., 2024). The
proliferation of sensor networks heightens risks associated with privacy
and surveillance; technical measures such as federated learning and
differential privacy are necessary but insufficient without ongoing
political and institutional oversight, as seen in prior failures in
algorithmic triage and automation of care workflows (The Guardian, 2021;
The Wall Street Journal, 2023). Historical underrepresentation in
training data perpetuates structural biases and exclusion, necessitating
equity audits, participatory metric design, and counterfactual analysis
to detect and remediate disparate impacts (Winner, 1980; Vidgen et al.,
2023). Dependence on highly interconnected AGI infrastructures increases
susceptibility to cyberattacks and cascading system failures; strategies
such as dual-mode fallback architectures and air-gapped control loops
can provide resilience (Amodei et al., 2016). Open-ended learning
dynamics present the possibility of unanticipated behaviors beyond
policy intent, making sandboxed simulations and phased deployment
important for safe evaluation of system responses (Sucholutsky et al.,
2023).

Scenario analysis enables structured exploration of possible urban
governance trajectories under AGI integration (Table 3). These include
five archetypal scenarios: participatory abundance, technocratic
efficiency, corporate enclaves, authoritarian panopticon, and stalled
transition. Each is defined by differences in oversight, resource
distribution, civic engagement, and ecological outcomes. Monitoring lead
indicators, including ownership concentration, audit transparency, and
participation rates, facilitates iterative adjustment of URL parameters
and supports adaptive governance.

In participatory abundance scenarios, democratic institutions and AI
systems operate under transparent auditing to distribute
automation-derived gains, foster civic engagement, and pursue
sustainable urban metabolism. Technocratic efficiency scenarios feature
managerial oversight and limited participation, resulting in baseline
welfare measures and moderate inequality. Corporate enclaves arise when
governance functions are assumed by private entities, resulting in
concentrated resources and fragmented public goods. Authoritarian
panopticon scenarios centralize allocation and surveillance, linking
distribution to compliance and loyalty. Stalled transitions reflect
regulatory inertia and limited progress in social or ecological domains.
These scenario constructs function as analytic tools for evaluating the
consequences of policy and design decisions and for calibrating
supervisory mechanisms such as the URL in pursuit of equitable and
sustainable outcomes.

Urban futures under AGI integration span diverse governance and ecological trajectories (Table~\ref{tab:scenarios}). Key indicators (ownership concentration, audit transparency, participation rates) serve as lead metrics for steering away from dystopian outcomes.

\begin{table}[ht]
\centering
\small
\renewcommand{\arraystretch}{1.25} 
\caption{Urban futures under AGI integration: governance, distribution, and ecological outcomes.}
\label{tab:scenarios}
\begin{tabularx}{\textwidth}{@{}%
    >{\raggedright\arraybackslash}X
    >{\raggedright\arraybackslash}X
    >{\raggedright\arraybackslash}X
    >{\raggedright\arraybackslash}X
    >{\raggedright\arraybackslash}X
@{}}
\toprule
\textbf{Scenario} & \textbf{Governance} & \textbf{Distribution} & \textbf{Socio-cultural outcome} & \textbf{Ecological outcome} \\
\midrule
Participatory Abundance  & Democratic–AI partnership; open audits          & Socialized AGI dividends                   & Robust civic culture; high volunteerism            & Net-zero metabolism                        \\
Technocratic Efficiency   & Managerial AI; limited oversight                & Moderate inequality; basic-income floor      & Paternalistic consumerism                          & Carbon-neutral; biodiversity loss          \\
Corporate Enclaves        & Platform oligopoly                              & Extreme inequality                          & Gated innovation zones                             & Green islands; neglected periphery         \\
Authoritarian Panopticon  & Central-state AI                                & Redistribution tied to loyalty               & Algorithmic policing                               & Eco-efficiency over justice                \\
Stalled Transition        & Regulatory gridlock                             & Persistent precarity                         & Erosion of trust                                   & Incremental emissions decline              \\
\bottomrule
\end{tabularx}
\end{table}

\noindent\textit{Interpreting Table~\ref{tab:scenarios}.} The five archetypes are deliberately schematic: they do not predict the future so much as map a spectrum of plausible trajectories against which policymakers can stress-test interventions. Each cell in the matrix highlights trade-offs between efficiency and equity, innovation and oversight, ecological ambition and political feasibility, rather than fixed outcomes. More severe or unanticipated hazards, including low-probability but high-impact existential risks such as irreversible value misalignment, global cascading failures, or runaway capability escalation, remain conceivable beyond the boundaries of these scenarios. Conversely, high rates of citizen participation, although correlated with closer alignment to local norms (Mushkani et al., 2025c), cannot by themselves guarantee that advanced systems will be free of harm. They mainly anchor decision criteria in place-based values and make any residual disagreements more transparent. The table should therefore be viewed as a living diagnostic tool whose categories and indicators require periodic revision as technical capabilities, social priorities, and risk perceptions evolve.

\section{Conclusion}
The integration of AI and prospective AGI into urban governance
constitutes a substantive transformation of cities as socio-technical
systems. Historical and comparative analysis indicates that, in the
absence of intentional oversight, automation processes tend to reinforce
existing inequalities and concentrate authority. The fragmentation and
limited enforceability of current regulatory approaches point to the
need for governance frameworks that embed normative standards,
participatory mechanisms, and adaptive oversight.

This study has articulated the URL as a
supervisory control structure, along with adjustable autonomy tiers and
participatory revenue-allocation models, for translating legal and
community standards into operational criteria for municipal AI systems.
The implementation of these frameworks depends on collaborative
engagement among policymakers, urban planners, technical developers, and
affected communities. Pilot projects in real-world environments enable
co-design, empirical evaluation, and incremental adaptation of both
technical and participatory protocols.

The limitations of the URL approach are acknowledged. Participatory
processes may be affected by representational bias and variable levels
of engagement. The technical complexity of AI systems may limit the
capacity of non-expert participants to engage with normative
threshold-setting. Risks of adversarial manipulation and the challenge
of scaling the URL across varied legal and institutional contexts remain
areas for further investigation. Addressing these limitations requires
ongoing empirical research, the iterative refinement of protocols, and
sustained interdisciplinary collaboration.

Further work should prioritize empirical validation in partnership with
municipalities, systematic comparison with existing governance models,
and the open sharing of protocols and datasets. Such cumulative
evidence-building can inform responsible integration of AI and AGI in
urban settings.

The trajectory of urban AI governance remains undetermined. Deliberate
engagement with the design and oversight of AI and AGI integration
provides the opportunity to align technological development with
normative commitments to equity, accountability, and sustainability. The
alternative is the continued emergence of opaque and inequitable
governance structures. The outcomes will depend on collective choices
and sustained institutional attention to these dynamics.

\section*{References}
{\small

Acemoglu, D., \& Restrepo, P. (2018). The race between man and machine:
Implications of technology for growth, factor shares, and employment.
American Economic Review, 108(6), 1488--1542.
\url{https://doi.org/10.1257/aer.20160696}

Alain, G., \& Bengio, Y. (2016). Understanding intermediate layers using
linear classifier probes (arXiv:1610.01644). arXiv.
https://doi.org/10.48550/arXiv.1610.01644

Altmeyer, P., Demetriou, A. M., Bartlett, A., \& Liem, C. C. S. (2024).
\emph{Position: Stop Making Unscientific AGI Performance Claims}
(arXiv:2402.03962). arXiv.
\url{https://doi.org/10.48550/arXiv.2402.03962}

Amodei, D., Olah, C., Steinhardt, J., Christiano, P., Schulman, J., \&
Mané, D. (2016). Concrete problems in AI safety. arXiv preprint
arXiv:1606.06565.

Arendt, Hannah \& Canovan, Margaret (1998). The Human Condition: Second
Edition. University of Chicago Press.

Arnstein, S. R. (1969). \emph{A Ladder of Citizen Participation}.
\emph{Journal of the American Institute of Planners}.

Attard-Frost, B. (2025). \emph{Transfeminist AI Governance}
(arXiv:2503.15682). arXiv. https://doi.org/10.48550/arXiv.2503.15682

Batty, M. (2018). \emph{Inventing Future Cities} (Illustrated edition).
The MIT Press.

Bibri, S. E., \& Krogstie, J. (2019). Generating a vision for smart
sustainable cities of the future: A scholarly backcasting approach.
\emph{European Journal of Futures Research}, \emph{7}(1), 5.
\url{https://doi.org/10.1186/s40309-019-0157-0}

Blyth v. Birmingham Waterworks Co., 11 Exch. 781, 156 Eng. Rep. 1047 (Exch. Ct. 1856).

Bostrom, N. (2014). Superintelligence: Paths, dangers, strategies.
Oxford University Press.

Bradford, A. (2023). Digital Empires: The Global Battle to Regulate
Technology. Oxford University Press.

Cardullo, P. (2020). \emph{Citizens in the ``Smart City'':
Participation, Co-production, Governance}. Routledge.
https://doi.org/10.4324/9780429438806

Castelfranchi, C., \& and Falcone, R. (2000). Trust and control: A
dialectic link. \emph{Applied Artificial Intelligence}, \emph{14}(8),
799--823. \url{https://doi.org/10.1080/08839510050127560}

Christian, B. (2020). The alignment problem: Machine learning and human
values. WW Norton \& Company.

Dai, J. (2024). \emph{Beyond Personhood: Agency, Accountability, and the
Limits of Anthropomorphic Ethical Analysis} (arXiv:2404.13861; Version
1). arXiv. https://doi.org/10.48550/arXiv.2404.13861

Deibel, T. U., \& Deibel, E. (2023). Artificial Intelligence in Ancient
Rome: Classical Roman Philosophy on Legal Subjectivity. In \emph{Women
Philosophers on Economics, Technology, Environment, and Gender History}
(pp. 157--168). De Gruyter.
https://www.degruyterbrill.com/document/doi/10.1515/9783111051802-017/html

European Commission. (2018). \emph{Ethics Guidelines for Trustworthy
AI}. \url{https://ec.europa.eu/futurium/en/ai-alliance-consultation}

Farnsworth, E. A. (1967). ``\emph{Meaning'' in the Law of Contracts.}
The Yale Law Journal, 76(5), 939--965.

Floridi, L. (1999). Information ethics: On the philosophical foundation
of computer ethics. \emph{Ethics and Information Technology}, 1(1),
33--52. \url{https://doi.org/10.1023/A:1010018611096}

Floridi, L. (2002). On the intrinsic value of information objects and
the infosphere. \emph{Ethics and Information Technology}, 4(4),
287--304. \url{https://doi.org/10.1023/A:1021342422699}

Gebru, T., \& Torres, É. P. (2024). The TESCREAL bundle: Eugenics and
the promise of utopia through artificial general intelligence.
\emph{First Monday}. \url{https://doi.org/10.5210/fm.v29i4.13636}

Geels, F. W. (2002). Technological transitions as evolutionary
reconfiguration processes: A multi-level perspective and a case study.
Research Policy, 31(8--9), 1257--1274.
\url{https://doi.org/10.1016/S0048-7333(02)00062-8}

Gunkel, D. J. (2006). The machine question: Ethics, alterity, and
technology. Explorations in Media Ecology, 5(4), 259--278.
\url{https://doi.org/10.1386/eme.5.4.259_1}

Gunkel, D. J. (2018). Robot rights. MIT Press.

Gurnee, W., \& Tegmark, M. (2023). Language models represent space and
time (arXiv:2310.02207v2). arXiv.
\url{https://doi.org/10.48550/arXiv.2310.02207}

Harris, J., \& Anthis, J. R. (2021). The moral consideration of
artificial entities: A literature review. Science and Engineering
Ethics, 27(4), 53. \url{https://doi.org/10.1007/s11948-021-00331-8}

Hughes, E., Dennis, M., Parker-Holder, J., Behbahani, F., Mavalankar,
A., \& Rocktaschel, T. (2024). Open-endedness is essential for
artificial superhuman intelligence. In Proceedings of the \emph{41st
International Conference on Machine Learning} (PMLR 235). Vienna,
Austria.

Kim, H., Yi, X., Yao, J., Lian, J., Huang, M., Duan, S., Bak, J., \&
Xie, X. (2024). The Road to Artificial Superintelligence: A
Comprehensive Survey of Superalignment.

Korinek, Anton. 2023. "Generative AI for Economic Research: Use Cases
and Implications for Economists." \emph{Journal of Economic Literature}
61 (4): 1281--1317.

Lefebvre, H. (1968). \emph{Le droit à la ville}.

Maslej, N., Fattorini, L., Perrault, R., Parli, V., Reuel, A.,
Brynjolfsson, E., Etchemendy, J., Ligett, K., Lyons, T., Manyika, J.,
Niebles, J. C., Shoham, Y., Wald, R., \& Clark, J. (2024).
\emph{Artificial Intelligence Index Report 2024} (arXiv:2405.19522).
arXiv. \url{https://doi.org/10.48550/arXiv.2405.19522}

Mei, Y., \& Robitscher, J. (2025). Illusion of Rights-Based AI
Regulation. Emory Law School.

Bradford, A. (2023). Digital Empires: The Global Battle to Regulate
Technology. Oxford University Press.

McKinsey Global Institute. (2017). A future that works: Automation,
employment, and productivity. McKinsey \& Company.

Morley, N. (2017, November 29). Return of the slave society. The Sphinx
(blog post).
\url{https://thesphinxblog.com/2017/11/29/return-of-the-slave-society/}

Mostafa, S. A., Ahmad, M. S., \& Mustapha, A. (2019). Adjustable
autonomy: A systematic literature review. \emph{Artificial Intelligence
Review}, \emph{51}(2), 149--186.
\url{https://doi.org/10.1007/s10462-017-9560-8}

Mushkani, R., Berard, H., Cohen, A., \& Koseki, S. (2025a). Position: The Right to AI. In Proceedings of the 42nd International Conference on Machine Learning (ICML 2025). https://doi.org/10.48550/arXiv.2501.17899

Mushkani, R., Berard, H., Ammar, T., Chatonnier, C., \& Koseki, S. (2025b). Co-Producing AI: Toward an Augmented, Participatory Lifecycle. In Proceedings of the 2025 AAAI/ACM Conference on AI, Ethics, and Society (AIES 2025). https://arxiv.org/abs/2508.00138

Mushkani, R., Nayak, S., Berard, H., Cohen, A., Koseki, S., \& Bertrand, H. (2025c). LIVS: A Pluralistic Alignment Dataset for Inclusive Public Spaces. In Proceedings of the 42nd International Conference on Machine Learning (ICML 2025). https://arxiv.org/abs/2503.01894

OpenAI. (2023). Governance of superintelligence. OpenAI.
\url{https://openai.com/index/governance-of-superintelligence/}

Pynadath, D. V., Scerri, P., \& Tambe, M. (2002). Towards Adjustable
Autonomy for the Real World. \emph{Journal of Artificial Intelligence
Research}, \emph{17}, 171--228. \url{https://doi.org/10.1613/jair.1037}

Petrov, A., La Malfa, E., Torr, P. H. S., \& Bibi, A. (2024). Prompting
a Pretrained Transformer Can Be a Universal Approximator. arXiv preprint
arXiv:2402.14753.

Qi, X., et al. (2023). Fine-tuning Aligned Language Models Compromises
Safety, Even When Users Do Not Intend to! arXiv preprint
arXiv:2310.03693.

Ramesh, A., Pavlov, M., Goh, G., Gray, S., Voss, C., Radford, A., Chen,
M., \& Sutskever, I. (2021). Zero-shot text-to-image generation. In
International Conference on Machine Learning (pp. 8821--8831). PMLR.

Rane, S. (2024). \emph{The Reasonable Person Standard for AI}
(arXiv:2406.04671). arXiv. https://doi.org/10.48550/arXiv.2406.04671

Sanchez, T. W., Fu, X., Yigitcanlar, T., \& Ye, X. (2024). The Research
Landscape of AI in Urban Planning: A Topic Analysis of the Literature
with ChatGPT. \emph{Urban Science}, \emph{8}(4), Article 4.
\url{https://doi.org/10.3390/urbansci8040197}

Schwartz, W. F. (1989). Objective and subjective standards of
negligence: Defining the reasonable person to induce optimal care and
optimal populations of injurers and victims. Georgetown Law Journal, 78,
241.

Sieber, R., Brandusescu, A., Adu-Daako, A., \& Sangiambut, S. (2024).
Who are the publics engaging in AI? \emph{Public Understanding of
Science}, \emph{33}(5), 634--653.
\url{https://doi.org/10.1177/09636625231219853}

Stiefenhofer, P. (2025). Artificial general intelligence and the end of
human employment: The need to renegotiate the social contract.
\emph{Department of Economics, Newcastle University}.

Sorensen, T., Moore, J., Fisher, J., Gordon, M., Mireshghallah, N.,
Rytting, C. M., Ye, A., Jiang, L., Lu, X., Dziri, N., Althoff, T., \&
Choi, Y. (2024). \emph{A Roadmap to Pluralistic Alignment}
(arXiv:2402.05070). arXiv.
\url{https://doi.org/10.48550/arXiv.2402.05070}

Solum, L. B. (1992). Legal personhood for artificial intelligences.
North Carolina Law Review, 70, 1231--1287.

Sucholutsky, I., Muttenthaler, L., Weller, A., Peng, A., Bobu, A., Kim,
B., Love, B. C., Grant, E., Achterberg, J., Tenenbaum, J. B., et al.
(2023). Getting aligned on representational alignment. arXiv preprint
arXiv:2310.13018.

Sun, H. (2020). Reinvigorating the human right to technology. Michigan
Journal of International Law, 41(2), 279.

Taylor, C. (1985). Philosophical Papers: Human Agency and Language.
Cambridge University Press.

The Guardian. (2021). What happened when a `wildly irrational' algorithm
made crucial healthcare decisions. Retrieved from
https://www.theguardian.com/us-news/2021/jul/02/algorithm-crucial-healthcare-decisions

The Wall Street Journal. (2023). When AI Overrules the Nurses Caring for
You. Retrieved from
https://www.wsj.com/articles/ai-medical-diagnosis-nurses-f881b0fe

Trager, R., Harack, B., Reuel, A., Carnegie, A., Heim, L., Ho, L.,
Kreps, S., Lall, R., Larter, O., hÉigeartaigh, S. Ó., Staffell, S., \&
Villalobos, J. J. (2023). \emph{International Governance of Civilian AI:
A Jurisdictional Certification Approach} (arXiv:2308.15514). arXiv.
\url{https://doi.org/10.48550/arXiv.2308.15514}

United States v. Carroll Towing Co., 159 F.2d 169 (2d Cir. 1947). 

Vaughan v. Menlove, 3 Bing. (N.C.) 468, 132 Eng. Rep. 490 (C.P. 1837).

Vidgen, B., et al. (2023). SimpleSafetyTests: A Test Suite for
Identifying Critical Safety Risks in Large Language Models. arXiv
preprint arXiv:2311.08370.

Wenar, L. (2023). Rights. In The Stanford Encyclopedia of Philosophy.

Winner, L. (1980). Do artifacts have politics? \emph{Daedalus}, 109(1),
121--136.

Wright, R. W. (2002). Justice and reasonable care in negligence law.
American Journal of Jurisprudence, 47, 143.

Zhang, P., Zhang, H., Xu, H., Xu, R., Wang, Z., Wang, C., Garg, A., Li,
Z., Ajoudani, A., \& Liu, X. (2025). Autonomous generalist scientist:
Towards and beyond human-level scientific research with agentic and
embodied AI and robots. DOI:10.13140/RG.2.2.35148.01923.

Zhang, J., Hua, X.-S., Huang, J., Shen, X., Chen, J., Zhou, Q., Fu, Z.,
\& Zhao, Y. (2019). City brain: Practice of large-scale artificial
intelligence in the real world. \emph{IET Smart Cities}, \emph{1}(1),
28--37. \url{https://doi.org/10.1049/iet-smc.2019.0034}
}

\end{document}